\documentclass[]{aastex631}

\newcommand{\ISOIS}{IS$\odot$IS}

\usepackage{multirow}
\usepackage{hyperref}
\usepackage{amsmath}
\usepackage{graphicx}
\usepackage{changepage}
\usepackage{soul}

\accepted{October 24, 2024}

\submitjournal{ApJL}

\begin{document}

\title{Composition variation of the May 16 2023 Solar Energetic Particle Event observed by Solar Orbiter and Parker Solar Probe}

\author[0000-0002-9246-996X]{Z. Xu}
\affiliation{California Institute of Technology, MC 290-17, Pasadena, CA 91125} \email{zgxu@caltech.edu}

\author[0000-0002-0978-8127]{C.M.S Cohen}
\affiliation{California Institute of Technology, MC 290-17, Pasadena, CA 91125}

\author[0000-0002-0156-2414]{R.A. Leske}
\affiliation{California Institute of Technology, MC 290-17, Pasadena, CA 91125}
\author[0000-0003-0581-1278]{G.D. Muro}
\affiliation{California Institute of Technology, MC 290-17, Pasadena, CA 91125}
\author[0000-0002-3840-7696]{A.C. Cummings}
\affiliation{California Institute of Technology, MC 290-17, Pasadena, CA 91125}
\author[0000-0001-6160-1158]{D.J. McComas}
\affiliation{Department of Astrophysical Sciences, Princeton University, Princeton, USA}
\author[0000-0002-3737-9283]{N.A. Schwadron}
\affiliation{University of New Hampshire, Durham, NH 03824, USA}
\author[0000-0003-2134-3937]{E.R. Christian}
\affiliation{Goddard Space Flight Center, Greenbelt, USA}
\author[0000-0002-2825-3128]{M.E. Wiedenbeck}
\affiliation{Jet Propulsion Laboratory, California Institute of Technology, Pasadena}
\author[0000-0002-4722-9166]{R.L. McNutt}
\affiliation{Johns Hopkins University Applied Physics Laboratory Laurel,  USA}
\author[0000-0003-1960-2119]{D.G. Mitchell}
\affiliation{Johns Hopkins University Applied Physics Laboratory Laurel,  USA}
\author[0000-0003-2169-9618]{G.M. Mason}
\affiliation{Johns Hopkins University Applied Physics Laboratory Laurel,  USA}
\author[0000-0001-6589-4509]{A. Kouloumvakos}
\affiliation{Johns Hopkins University Applied Physics Laboratory Laurel,  USA}
\author[0000-0002-7388-173X]{R.F. Wimmer-Schweingruber}
\affiliation{Institute of Experimental and Applied Physics, Kiel University, Germany}
\author[0000-0003-1093-2066]{G.C. Ho}
\affiliation{Southwest Research Institute, San Antonio, TX 78228, USA}
\author[0000-0002-4240-1115]{J. Rodriguez-Pacheco}
\affiliation{University of Alcala, Space Research Group, Alcalá de Henares, Madrid, Spain }



\begin{abstract}
In this study, we employ the combined charged particle measurements from Integrated Science Investigation of the Sun (\ISOIS) onboard the Parker Solar Probe (PSP) and Energetic Particle Detector (EPD) onboard the Solar Orbiter (SolO) to study the composition variation of the solar energetic particle (SEP) event occurring on May 16, 2023. During the event, SolO and PSP were located at a similar radial distance of $\sim$0.7 au and were separated by $\sim$60$^\circ$ in longitude. The footpoints of both PSP and SolO were west of the flare region but the former was much closer (18$^\circ$ vs 80$^\circ$). %\st{Such an ideal distribution of observers allows us to study} 
Such a distribution of observers is ideal for studying the longitudinal dependence of the ion composition with the minimum transport effects of particles along the radial direction. We focus on H, He, O, and Fe measured by both spacecraft in sunward and anti-sunward directions. Their spectra are in a double power-law shape, which is fitted best by the Band function. Notably, the event was Fe-rich at PSP, where the mean Fe/O ratio at energies of 0.1 - 10 Mev/nuc was 0.48, higher than the average Fe/O ratio in previous large SEP events. In contrast, the mean Fe/O ratio at SolO over the same energy range was considerable lower at 0.08. The Fe/O ratio between 0.5 and 10 MeV/nuc at both spacecraft is nearly constant. Although the He/H ratio shows energy dependence, decreasing with increasing energy, the He/H ratio at PSP is still about twice as high as that at SolO.
Such a strong longitudinal dependence of element abundances and the Fe-rich component in the PSP data could be attributed to the direct flare contribution.
%or the flare remnant particles accelerated in the vicinity of the nose area by the shock driven by coronal mass ejection.
Moreover, the temporal profiles indicate that differences in the Fe/O and He/H ratios between PSP and SolO persisted throughout the entire event rather than only at the start. 
%Thus, the observational discrepancy may also partially be ascribed to transport effects.

\end{abstract}



\section{Introduction} \label{sec:intro}

Solar energetic particle (SEP) events represent a significant threat to astronauts and electronic devices onboard spacecraft during space exploration missions. With the increasing number of human space activities in deep space and on planetary surfaces, the demand to understand the generation, variability, and radiation effects of SEPs is becoming increasingly critical. SEPs are accelerated by two processes related to fierce solar eruptions, the flare-related magnetic reconnection process and diffusive shock acceleration driven by coronal mass ejections (CMEs) \citep{Reames1999SSRv,Kallenrode2003JPhG, Cane2003GeoRL}. The latter usually causes large SEP events with extremely high intensity, long duration, and fairly wide longitudinal spread within the heliosphere.

The variation in the composition of large SEP events has long been a question of investigations. The difficulty in fully understanding those variations arises primarily because the characteristics of energetic particle events observed by distant observers, such as those near the first Lagrange point (L1) point, are the combined result of the acceleration mechanism and transport effects within the interplanetary medium \citep{Cohen2017ApJ}.
%The reason is due to the complicated acceleration mechanism and transport effect.
One surprising observation is the enhancement of the Fe/O ratio in large SEP events, often referred to as Fe-rich events \citep{Cohen1999GeoRL, Cohen1999GeoRL_New, Mason1999GeoRL}. Currently, the causes of these Fe-rich events, and how the Fe/O ratio varies as a function of energy, are still under debate \citep{reames1998SSRv,Cane2003GeoRL, Cohen2005JGR,Tylka2005ApJ, Desai2016LRSP}.  This is partly due to the sparse spatial distribution of in-situ measurements in the heliosphere.
%Then you can say that we still don't really have agreement on this.

From previous studies, two proposed scenarios have emerged to explain the enhancement of the heavy ions during these large events. The first is the direct flare contribution \citep{Cane2003GeoRL, Cane2006JGR}, where Fe-rich material from flare-related acceleration is present along with shock accelerated material.  %The second relies on perpendicular shock acceleration of a flare-related suprathermal seed population \citep{Tylka2005ApJ}.  
The second relies on the acceleration of a flare-related suprathermal seed population at shock waves with a perpendicular shock geometry \citep{Tylka2005ApJ}. Additionally, transport effects have been suggested to explain the temporal variability of Fe/O, specifically the high Fe/O ratio at the start of an SEP event \citep{Mason2006ApJ, MasonInterplantary2012ApJ, Mason2014ASPC, Dalla2017AA}.

An effective method to differentiate and test the initial two scenarios is to examine the longitudinal dependence of the Fe/O ratio \citep{Cohen2014ApJ, Cohen2017ApJ}. Should the Fe/O ratio exhibit significant longitudinal dependence, with an observer magnetically well connected to the flaring region recording higher values than a poorly connected observer, it would support the flare direct contribution hypothesis. In contrast, the shock acceleration scenario would likely exhibit a completely limited or no longitudinal variation due to the locality of the perpendicular shock and the flare suprathermal seed particles. Meanwhile, the transport effect can be studied through examining  the time variability of the composition for all the observers and particularly benefits from observations at different radial distances.  Given the need to utilize  multi-spacecraft measurements of Fe/O over a broad energy range, and potentially from locations closer to the Sun where the transport effects are less significant, opportunities to obtain a clear answer to the cause of the elevated Fe/O abundance ratios in SEP events have been few.

Solar Orbiter (SolO) mission \citep{Mueller-2020-SolO} was launched on February 10, 2020 at the start of Solar Cycle (SC) 25. Through a series of Venus flybys, SolO has attained a close perihelion distance of around 0.3 au. Parker Solar Probe (PSP) \citep{Fox2016SSRv}, launched earlier on August 12, 2018, has gone through 20 orbits and will soon achieve a closest perihelion distance of less than 10 solar radii (0.046 au).
Taking advantage of their high-quality charged particle data measured at these unprecedented locations in the inner heliosphere, PSP and SolO provide more opportunities to study the longitudinal and radial dependence of  SEP composition variability, and hopefully shed light on the causes of the enhanced Fe/O abundances.
A prime opportunity for this was the 16 May 2023 SEP event.  The event was observed by SolO and PSP when both were at similar distances of approximately 0.7 AU but were significantly separated in longitude ($>60^\circ$).  PSP was magnetically well connected to the flaring region while SolO's magnetic footpoint  was farther to the west. Such a configuration is perfect for examining the longitudinal dependence of the Fe/O ratio.

\section{Observations}

\subsection{Instrumentation}
In this study, observations of energetic H, $^4$He, O, and Fe ions are provided by the Energetic Particle Detector (EPD, \cite{RodriguezPacheco-2019-EPD, Wimmer2021AA}) instrument suite onboard SolO, and Integrated Science Investigation of the Sun (\ISOIS, \cite{McComas2016SSRv_ISOIS}) onboard PSP.
The SolO protons with energies ranging from tens of keV to hundreds of MeV are measured by the High Energetic Telescope (HET) and the Electron-Proton Telescope (EPT). SolO/HET can also measure and distinguish different ion species by employing the dE/dx versus total energy technique. The energy range is from several MeV/nuc to a few hundred MeV/nuc. Unfortunately, during this SEP event, at energies above 10 MeV/nuc the O and Fe intensities were too low to be measured by SolO/HET.

%SolO/HET was only able to measure $^4$He and H.

%hardly measured enough number of heavy elements O and Fe above 10 MeV/nuc due to their lower intensity.
%Not to speak of that SolO was not situated a well-connected location related to the flare region (Below for more details).
%unlike those SEP events occurred in the previous solar cycle\cite{}, even though SolO and PSP are much closer to the Sun than those L1 missions. 
%Heavy ions 
He, O, and Fe with energies between hundreds keV/nuc and few MeV/nuc are measured by the Suprathermal Ion Spectrograph (SIS), which is the time-of-flight mass spectrometer of EPD. Both SIS, EPT, and SolO/HET have multiple apertures facing different directions; only measurements from the sunward and anti-sunward telescopes are used in this study.

The corresponding SEP measurements from PSP/\ISOIS \ are obtained by low-energy (EPI-Lo) and high-energy (EPI-Hi) energetic particle instruments \citep{Hill2017JGRA, Wiedenbeck2017ICRC}. EPI-Lo is a time-of-flight mass spectrometer consisting of 80 apertures covering 2 $\pi$ field of view and measuring particles from $\sim$tens of keV to a few MeV.
%Note that the $^4$He intensities of EPI-Lo are corrected to match the intensities in the top apertures for the purpose of correcting the efficiencies of the EPI-Lo wedges. 
%\TODO{[I need several sentences to explin the correction of EpiLo He4, ask Don; and ask him provide opinion on the Proton spectra whether good  or not]}
%, which is the lower energy portion of the energetic particles spectra above the solar wind energy.%\cite{I need the Melweldt paper of the particle spectra}
EPI-Hi measures the higher energy portion of the \ISOIS \ energy coverage ($>$ few MeV/nuc). EPI-Hi consists of three Low Energy Telescopes (LET-A/B/C) and a pair of High Energy Telescopes (HET-A and HET-B). PSP/LET and PSP/HET utilize the same measurement principle as SolO/HET and SolO/EPT, i.e. the dE/dx versus total energy technique, and can differentiate elements from H to Fe. 

Here we use the sunward facing telescopes, LET-A and HET-A, and the anti-sunward facing telescopes, LET-B and HET-B to measure H, He, O, and Fe. However, similar to SolO, the intensities of O and Fe were not high enough to be detected by HET-A and HET-B. For EPI-Lo we have selected and averaged over those apertures that approximately correspond to the fields of view of LET-A\footnote{aperture numbers 21, 22, 23, 24, 25, 26, 27, 28, 29, 30, 32, 34, 35, 36, 37, 38, 39, 44, 46, 48} and LET-B\footnote{aperture numbers 61, 62, 63, 70, 71, 72, 73}.

The Solar Orbiter magnetometer (MAG, \cite{Horbury2020AA}) and the FIELDS instrument suite \citep{Bale2016SSRv_FIELD} of PSP provide direct measurements of the magnetic field. Apart from the in situ measurements, the remote-sensing observations from the white-light coronagraph onboard SOHO, Large Angle and Spectrometric Coronagraph (LASCO, \cite{Brueckner1995SoPh}) are used to obtain information of the CME. 

%PSP/LET and PSP/HET measurements from the sunward (A) and anti-sunward (B) telescopes are employed later to ensure compatible observations between SolO and PSP. 
%PSP/LET provides measurements of all four types of elements, while PSP/HET additionally offers measurements of higher energy H and $^4$He. It is worth noting that the intensities of EPi-Lo sunward and antisunward are the mean flux of these small apertures within the field of view (FOV) of PSP/LET-A and PSP/LET-B.

%from other instruments including t and the Spectrometer Telescope for Imaging X-rays (STIX) \citep{SolOSTIX2020A&A}, as well as . These observations are crucial to understand the overall behaviors of this event.

\subsection{Overview of the May 16, 2023 event}

\begin{figure}
    \centering
    \includegraphics[width = \textwidth]{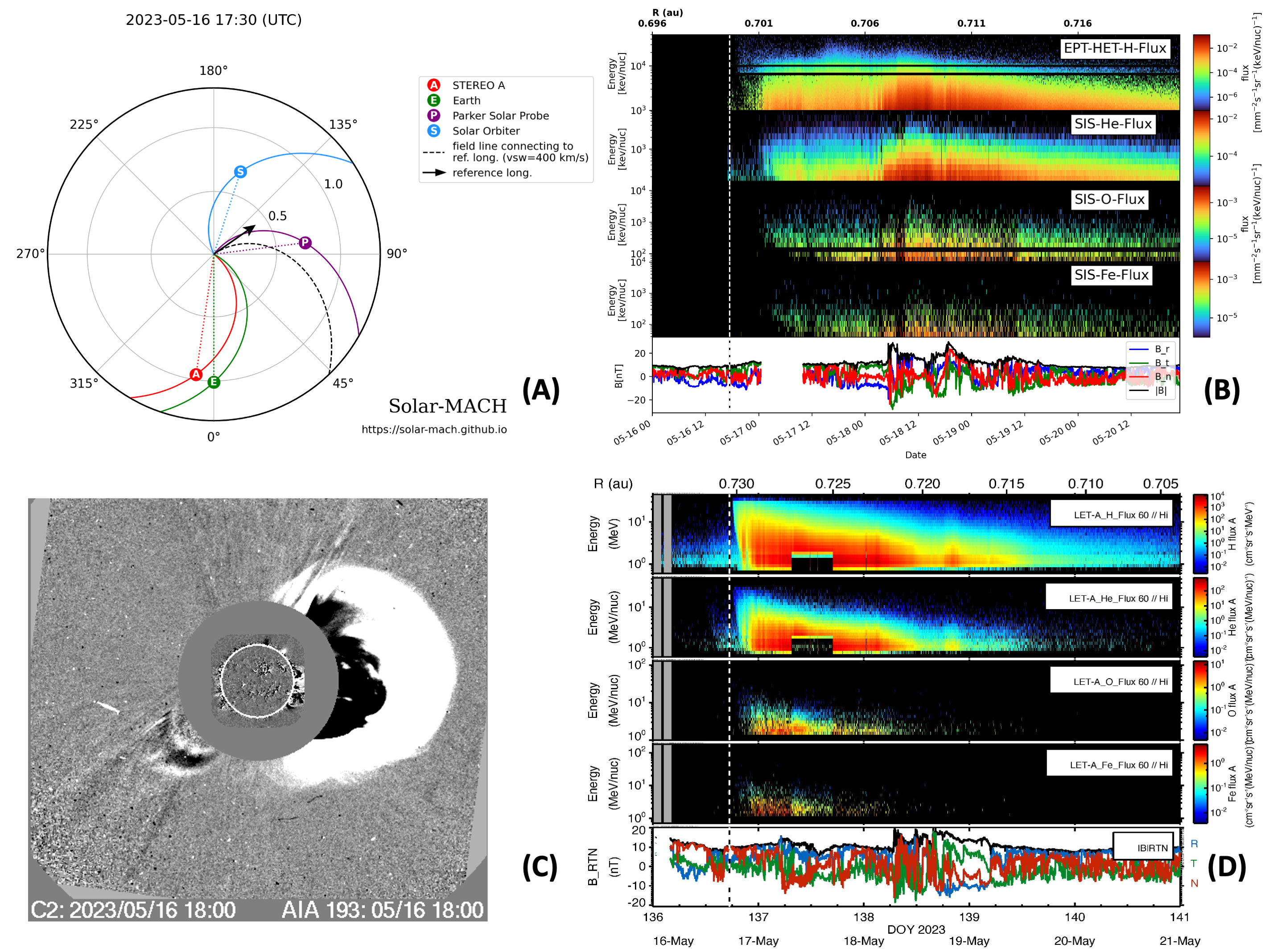}
    \caption{(A): The constellation of different spacecraft at the start of the SEP event, created by Solar Mach \citep{Gieseler2023FrASS}. The black arrow indicates the approximate longitudinal location of the flare; (B): The in-situ observations of energetic particles from SolO/EPD SIS, EPT, and HET. From top to bottom, the panels are the dynamic spectra of protons, helium, oxygen, and iron. The bottom panel displays the magnetic field magnitude and its components;  (C): The running difference image of the CME observed by SOHO/Lasco C2 at 18:00 on May 16, 2023; (D): The in-situ PSP/\ISOIS \ LET-A
    observations presented in the same format as panel (B). Note that the top axis of (B) and (D) indicates the radial distance of SolO and PSP from the Sun.} 
    \label{fig:Overview}
\end{figure}

The SEP event of 16 May 2023 was related to solar activity originating from a far-side (as viewed from Earth) active region, AR13296, located at (125W, 12.5N) \footnote{Stonyhurst coordinate system, determined by the Extreme Ultraviolet Imager (EUI, \citep{SolOEUI2020A&A}) onboard SolO}. In panel (A) of Fig.~\ref{fig:Overview}, the location of the flare is indicated by the black arrow. 
Different observers are displayed in the figure, including Earth (green), STEREO-A (red), PSP (purple), and SolO (blue). Magnetic connections between the Sun and different observers are drawn as the nominal Parker spiral lines which are calculated assuming a 400 km/s solar wind speed. As exhibited in panel (A), the magnetic footpoint of SolO lies approximately 80$^\circ$ to the west of the flare, while the footpoint of PSP is much closer, only 18$^\circ$ west of the active region. PSP and SolO have very similar radial distances, $\sim$0.72 au, and $\sim$0.70 au away from the Sun, respectively, suggesting observed differences in SEPs are not likely due to radial effects. 
%The Earth's magnetic foot point is connected to the east of the flare, at approximately 62 degrees.
%Notably, BepiColombo is situated about 0.58 AU from the Sun, with its magnetic footpoint approximately 50 degrees east of the activity region. Due to its closer proximity to the Sun compared to PSP and SolO, and the unique longitudinal distribution in interplanetary space, the corresponding SEP observations during this event should add complexity and interest. 
Apart from SolO, PSP and STEREO-A, instruments near the L1 point, such as the Electron Proton Helium Instrument (EPHIN, \cite{COSTEP1995SoPh}) onboard SOHO, Solar Isotope Spectrometer (SIS, \cite{Stone1998SSRv}) onboard ACE and the Lunar Lander Neutron \& Dosimetry Experiment (LND, \cite{Wimmer2020SSRv}) on the lunar far-side surface when LND was operating on the local day time \citep{Xu2020ApJ}, also recorded a noticeable increase in protons below 50 MeV, helium with energy below 10 MeV/nuc, and heavy ions in the range of a few hundred keV/nuc. The SEP event was observed by all these observers over a nearly 180$^\circ$ longitudinal extent, indicating the wide-spread nature of the event. In the following analysis, we will focus solely on energetic particle measurements from SolO and PSP.

%Before this event, AR13296 had multiple eruptive activities on May 8 when it was still facing the Earth. The early morning eruption even caused larger SEPs to reach Earth, as indicated by measurements from L1 instruments. 
% we present the different observers' position in space and their magnetic connection to the flare and active region on the solar surface .

Observations from the Spectrometer Telescope for Imaging X-Rays (STIX, \cite{SolOSTIX2020A&A}) onboard SolO show that the x-ray emissions reached their peak at 17:24 on 16 May 2023. 
%Note that the times mentioned in the context, including the peak time and the following times, have been adjusted for the light travel time which is 8.3 minutes at Earth and 5.8 minutes at PSP and SolO. 
Estimations from the STIX observations suggests that the corresponding GOES flare class was about M4.9. According to the PSP/FIELDS radio dynamic spectrum in the two bottom panels of Fig.\ref{Fig:radio}, the eruption was accompanied by multiple type III radio bursts between 17:16 and 17:24 and a type II radio burst,
%with lower than normal flux intensity,
%very low flux density, 
commonly considered as a proxy of a shock wave in the solar corona or IP space, starting from 17:24.
%of a CME-driven shock, together with its corresponding second harmonic wave structure starting from 17:24. 
A second group of type III radio bursts starts at about 17:34. 
%The detailed relationship between the radio emission and SEPs are out of the scope of this paper.
The associated CME and shock was first observed by SOHO/LASCO C2 at 17:36. It then traveled towards the PSP at a plane-of-sky speed of 1099 km/s with an angular width of 360\textdegree, as estimated from the SOHO LASCO CME catalog \footnote{\url{https://cdaw.gsfc.nasa.gov/CME_list/index.html}}\citep{Gopalswamy2009EM&P}.
%From the PSP's perspective, this is a halo CME. 
In panel (C) of Fig.~\ref{fig:Overview}, the SOHO/LASCO C2 running difference image shows the brightened CME front along with its extensive angular width. 

Panels (B) and (D) on the right side of Fig.~\ref{fig:Overview} display the dynamic spectrograms of  H, He, O, and Fe (from top to bottom) measured by SolO/EPD and PSP/\ISOIS \ from their sunward telescopes. The antisunward measurements are not shown but are utilized in subsequent sections. 
The vertical dashed lines indicate the time of the peak of the SolO/STIX SXR light curves at around 17:18, which includes the light travel time from the Sun to SolO.
%at around 17:18. 
The radial distances of SolO and PSP are given on the top axes of panel (B) and panel (D).

The peak proton intensity on PSP occurred on 16 May 2023 at $>$ 30 $cm^{-2} sr^{-1} MeV ^{-1} s^{-1}$ for energies between 10 - 25 MeV and the event lasted for about 5 days before intensities returned to pre-event levels. Whereas SolO recorded a much lower peak intensity of $\sim$1.5 $cm^{-2} sr^{-1} MeV ^{-1} s^{-1}$ in the same energy range and significantly later in the event, on 18 May 2023.
The SolO observations in panel (B) are typical of an SEP event originating from a CME-driven shock emanating from a source east of the observer \citep{Cane1988}. The intensity levels gradually increased, reaching their peak when the shock crossed the spacecraft on the morning of May 18. PSP measures a more rapid onset, with the intensities peaking early in the event, followed by a prolonged decay. The empty regions in the H and $^4$He panels between May 17 and 18 (at energies below 2 MeV/nuc) are due to the change in the dynamic threshold mode of PSP/LET. PSP/LET-A and PSP/LET-B raised the thresholds on some detectors to maintain a high instrument live time \citep{Wiedenbeck2017ICRC, Cohen2021AA} and these lower energies are not measured during this time. For this reason, the channels with energies below 2 MeV/nuc are not included in the subsequent spectral analysis. 

The proton measurements on both spacecraft exhibit velocity dispersion during the onset of the event, but the dispersion at SolO is not as clear as that at PSP. We present the velocity dispersion analysis (VDA) results for both cases (in table \ref{tab:location}). For PSP, the analysis relies primarily on PSP/LET-A measurements which have the highest intensity compared to other apertures, revealing a particle travel length of 0.95 $\pm$ 0.03 au and a particle release time of 17:31 $\pm$ 2 min. As for SolO/EPD, the velocity dispersion of high energy protons and lower energy protons cannot be described by one straight line, though both show clear velocity dispersion. The VDA from the higher energy particles produces a pathlength of 1.53 $\pm$ 0.15 au, with a release time of 17:34 $\pm$ 11 min, coincident with the second type III radio burst, and also consistent with the release time of the PSP protons. In contrast, the SolO protons with energies roughly less than 500 keV are released approximately 2.3 hours later with a travel length of 0.87 $\pm$ 0.06 au. A warning here is the lower energy proton might scatter more and violate the scatter-free assumption of the VDA method.
These two groups of particles have characteristics very similar to the two classes of proton events, as reported by \cite{Krucker2000ApJ}. Lower energy protons with shorter path lengths and late release times might be due to a later release after being accelerated by a coronal shock at a high altitude \citep{AKouloumvakos2023}. 

%Of course, VDA method relies on the assumption that all energy particles release simultaneously at the same time from the same site and that the propagation of those energetic particles is scattering-free. However, the existence of perpendicular diffusion in the heliospheric space and the unclear pre-event background could significantly affect the VDA result \citep{Wang2015ApJ}. Considering the larger magnetic separation between the flare and the foot point of SolO, the uncommon VDA fitting results of the different energy particles of SolO might be acceptable.

%The new studies  though the particle traveling paths are still worth further studies considering the new observation of the extremely longer particle propagation lengths that are affected by the interplanetary environment. \cite{wimmer2023}
% The extremely longer particle traveling length from the 9 APril 2022 event indicates the large scale magnetic structures can significantly affect the length.
At both locations, the arrival of the shock plus CME structure at the spacecraft is evidenced by the discontinuity in the total magnitude of B (black line) and the change in sign of the radial and tangential components, as depicted in the bottom panels of (B) and (D) of Fig.~\ref{fig:Overview}. %The CME reached SolO approximately 1.75 hours earlier than PSP. 

%On the one hand, SolO is slightly closer to the Sun than PSP. Such time shift could be accouted; On the other hand, is the flank speed and nose speed same? or the flank faster?
%Instruments near the L1 point, such as the Electron Proton Helium Instrument (EPHIN) \citep{COSTEP1995SoPh} onboard SOHO, Solar Isotope Spectrometer (SIS)\citep{Stone1998SSRv} onboard ACE and the Lunar Lander Neutron \& Dosimetry Experiment \citep{Wimmer2020SSRv} on the lunar far-side surface, also recorded a noticeable increase in protons below 50 MeV, helium with energy below 10 MeV, and heavy ions in the range of a few hundred keV/nuc, highlighting the widespread nature of SEP. 

\begin{table}[]
    \centering
    \begin{tabular}{|c|c|c|}
    \hline
          & \multicolumn{2}{c|}{Spacecraft}\\
          \hline
          & PSP & SolO \\
         \hline
         r (au)& 0.73 & 0.70 \\
         \hline
         lon$_{\rm stonyhurst}$  & 97.1  & 162.0 \\
         \hline
         lon$_{\rm footpoint}$ & 142.6  & 204.7 \\
         \hline
         D$_{\rm lon}$  & 17.6 & 79.7 \\
         \hline
         L$_{\rm VDA}$ (au)  & 0.95$\pm$0.03 & 1.53 $\pm$ 0.15 \\
         \hline
         T$_{\rm release}$  & 17:31 $\pm$3  & 17:34 $\pm$ 11  \\
         \hline
         Fe/O(0.1- 10 MeV/nuc)  & 0.48 $\pm$0.014 & 0.08 $\pm$ 0.01\\
         \hline
    \end{tabular}
    \caption{From top to bottom: The radial distance (r) and longitude (lon$_{\rm stonyhurst}$) of PSP and SolO; the longitude of magnetic footpoint (lon$_{\rm footpoint}$); the longitudinal separation (D$_{\rm lon}$) between the magnetic footpoint of PSP and SolO, and the location of the flare, which was at 125 degrees Stonyhurst longitude; the proton travel path derived from VDA (L$_{\rm VDA}$) and the estimated particle release time of all energy (T$_{\rm release}$); the averaged Fe/O ratio in the energy range of 0.1 to 10 MeV/nuc.}
    \label{tab:location}
\end{table}

%begin{table}[]
%    \centering
%    \begin{tabular}{c|c|c|c}
%    \hline
%         Spacecraft & Flare & PSP  & SolO\\
%         \hline
%         r (au) & -& 0.73 & 0.70 \\
%         \hline
%         lon$_{stonyhurst}$ & 125  & 97.1  & 162.0 \\
%         \hline
%         lon$_{footpoint}$ & -& 142.6  & 204.7 \\
%         \hline
%         D$_{lon}$ &- & 17.6 & 79.7 \\
%         \hline
%         L$_{VDA}$ (au) & - & 0.95$\pm$0.03 & 1.53 $\pm$ 0.15 \\
%         \hline
%         T$_{release}$ & - & 17:31 $\pm$3  & 17:34 $\pm$ 11  \\
%         \hline
%         Fe/O(0.1- 10 MeV/nuc) & - & 0.48 & 0.08 \\
%         \hline
%    \end{tabular}
%    \caption{From top to bottom: The radial distance and longitude of PSP and SolO; the longitude of magnetic footpoint; the longitudinal separation between the magnetic footpoint of PSP and SolO, and the location of the flare; the proton travel path derived from VDA and the estimated particle release time; the averaged Fe/O ratio in the energy range of 0.1 to 10 MeV/nuc.}
%    \label{tab:location}
%end{table}

\subsection{Integrated spectra}

Fig.~\ref{fig:spectra} presents the event-integrated spectra of  H (purple), He (green), O (orange), and Fe (blue) measured by both PSP and SolO. 
The spectra obtained from the sunward and anti-sunward telescopes are shown in separate panels for clarity.  
All the spectra depicted in Figure~\ref{fig:spectra} exhibit a characteristic bending-over from lower to higher energies. 
We have fitted the spectra using different spectral functions and find that the Band function, as proposed by \cite{Band1993ApJ}, best describes these spectra, yielding the minimal residual values. The Band function (see Equation 1 of \cite{Band1993ApJ} for details) can be expressed as
\begin{equation}
    F(E) = \begin{cases}
    A_0E^{-\beta_1}e^{-E/E_0},  &                                             
              E < (\beta_2 - \beta_1)E_0\\
    A_0E^{-\beta_2}[(\beta_2 - \beta_1)E_0]^{(\beta_2 - \beta_1)}e^{\beta_1-\beta_2}, & E > (\beta_2 - \beta_1)E_0 
  \end{cases}
\end{equation}
which is defined by four parameters, $A_0$, $\beta_1$, $\beta_2$, $E_0$. F(E) is the fluence of different elements. $\beta_{1}$ and $\beta_{2}$ represent the power law indices of the spectra at lower and higher energy ranges, respectively. The break energy equals $(\beta_2 - \beta_1) * E_0$ locates the transition between the lower and higher energy power law segments. A$_0$ is a constant parameter.

The corresponding fitted parameters are given in Table~\ref{tab:fitting_parameter}. 
Overall, the spectra steepen with increasing energy. In the high-energy range, the spectra indices ($\beta_2$) are between 2.68 and 5.19, with a mean value of 3.78. The lower energy spectra have a mean index value of 1.08. These values are consistent with the values obtained in the survey of large SEP events fitted with the Band function \citep{Desai2016ApJ}, though their survey only includes the SEPs without significant increases in intensity during the local shock passage. 
In addition, the differences between the sunward and anti-sunward spectra are minimal at both spacecraft, suggesting limited anisotropy for the majority of the event.
%The event-integrated spectra of SolO tend to be softened in comparison to PSP at both high-energy and low-energy. The transit energies $E_0$ of SolO are smaller than those of PSP. 
It is worth noting that the fitting parameters are strongly affected by the incomplete energy coverage of the spectra, especially the Fe spectra. For example, the Fe spectra of PSP have a broad data gap between the EPI-Lo data points (empty circles) and the EPI-Hi data points (filled circles), as shown in the bottom two panels of Fig.~\ref{fig:spectra}. The O and Fe spectra of SolO lack high energy measurement due to their low intensity. Consequently, their positions of the break energy and power-law indices in the high energy range above the break energy are not well determined and we do not include them  in  Table ~\ref{tab:fitting_parameter}.
%We hope that future events with higher intensities over a wider energy range 
%First, as discussed in the instrument section, EpiLo He spectra require additional calibrations, and directional normalization has been applied to the helium-4 measurements on different wedges of EPiLo to correct the detection efficiency. At present, the normalization process is applied exclusively to $^4$ He, and it is not taken into account for the other three elements. The potential impact of the detection efficiency issue on these other particle types is still being investigated.

%The most power spectra indices above the break energy above 3.
%The 
%The index xFe in SolO has large errorbar due to the incomplete coverage of the whole spectra.

%\todo[inline]{You might need to read several spectra paper, target on how they describe the spectra of particle, the last step; Check the parameter and ask Christina. Read some Desai's spectra paper. Of course the focus are not this part. I maybe just left it there.}

%can be expressed as
%\begin{equation} \label{Eq:1}
%\begin{split}
%f(E) &= A0*E^{-\beta_1} * exp(-E/E_0) ; E < E_{break} \\ 
%f(E) &= A0*E^{-\beta_2} * E_{break} ^{\beta_2-\beta_1} * exp(\beta_1 -\beta_2);  E > E_{break} \\
%E_{break} & = (\beta_2 - \beta_1) * E_0;
%\end{split}
%\end{equation}

%The E-R function not fitted: reason: might be due to the limited higher energy tail? Find the result do spectra analysis and find some information: \TODO{TODO here}

%The break energy is associated with the shock acceleration[?]

\subsection{Composition variation with energy} 

\begin{figure}
    \centering
    \includegraphics[width=\textwidth]{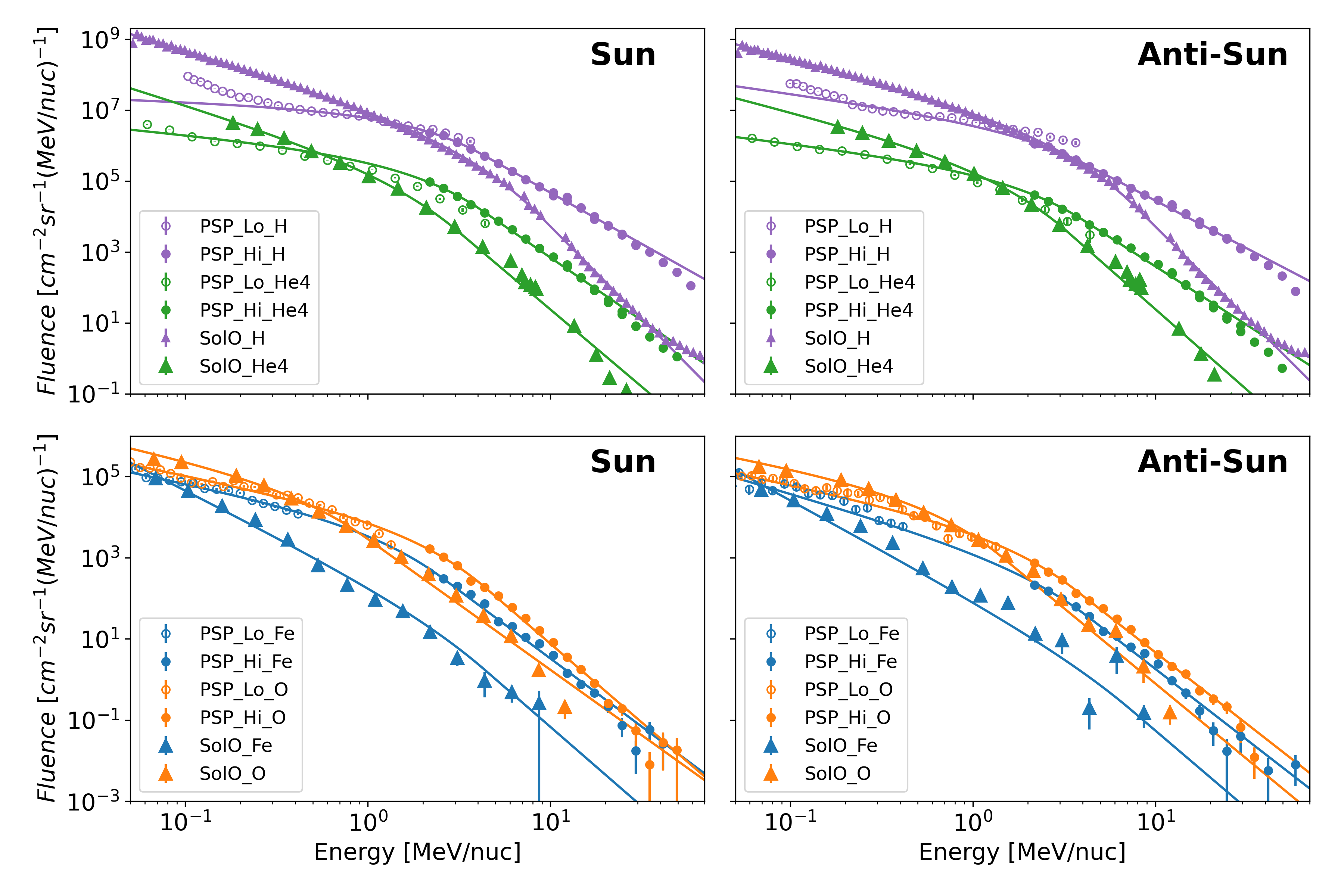}
    \caption{The fluence spectra of H (purple), $^4$He (green), O (orange), and Fe (blue) integrated between 2023-05-16T17:00:00 and 2023-05-21T00:00:00. Spectra from SolO are shown as triangles,  PSP/EPI-Lo as empty circles and PSP/EPI-Hi as filled circles. Two panels in the left half present the spectra measured by the sunward telescopes and the right half represents the spectra measured by the anti-sunward telescopes. The solid lines represent the best-fitted Band function \citep{Band1993ApJ}.} 
    \label{fig:spectra}
\end{figure}

\begin{table}
\begin{adjustwidth}{-3cm}{}
\begin{tabular}{|C|C|C|C|C|C|C|C|C|}
    \hline
    &\multicolumn{4}{c|}{Index($\beta$1, $\beta$2)} &\multicolumn{4}{C|}{E$_{0}$} \\
    \hline
        &  H & He & O & Fe & H & He & O & Fe\\
    \hline
     %PSP  & 0.19 $\pm$0.31   &1.18$\pm$0.03  &0.64$\pm$0.07  &0.74$\pm$ 0.06    &  &   &   & \\
    PSP  & 0.19 $\pm$0.31   &0.50$\pm$0.07  &0.90$\pm$0.04  &0.90$\pm$ 0.05    &  &   &   & \\
    % (sun)& 2.88 $\pm$ 0.03  &3.54$\pm$0.04  &3.57$\pm$0.06  &3.31$\pm$ 0.07    & 1.56 $\pm$ 0.26       &  1.84 $\pm$ 0.08 & 0.59 $\pm$ 0.05   & 0.45 $\pm$ 0.04  \\
    (sun)& 2.88 $\pm$ 0.03  &3.54$\pm$0.06  &3.87$\pm$0.12  &3.35$\pm$ 0.07    & 1.56 $\pm$ 0.26       &  1.38 $\pm$ 0.08 & 1.47 $\pm$ 0.08   & 1.03$\pm$ 0.08  \\
     \hline
     PSP   & 0.71$\pm$0.25, & 0.62 $\pm$ 0.11, & 0.94$\pm$ 0.05 & 1.30$\pm$0.07  &      & & &  \\
     (antisun)& 2.68 $\pm$0.03 & 3.30 $\pm$ 0.05  & 3.51$\pm$ 0.08 & 3.48$\pm$0.17  & 2.16$\pm$0.40        & 1.49 $\pm$ 0.11 & 1.40$\pm$ 0.11    &2.00 $\pm$ 0.29 \\
     \hline
     SolO & 1.57$\pm$ 0.01  & 1.59 $\pm$ 0.08   &0.98 $\pm$ 0.15 & 2.27 $\pm$ 0.15 & & & &  \\
     (sun)& 5.19 $\pm$ 0.32 & 4.35 $\pm$ 0.34   &3.22 $\pm$ 0.21 & ... & 2.37 $\pm$ 0.07    & 1.13 $\pm$ 0.17 & 0.43 $\pm$0.09  & ...\\
     %2.67 $\pm$ 2.48 \\
     %3.91 $\pm$ 8.01
     \hline
     SolO   & 1.36 $\pm$ 0.01  & 1.30 $\pm$ 0.05    &0.83$\pm$ 0.15 &2.45 $\pm$ 0.32 & & & &  \\
     (antisun) &5.13 $\pm$ 0.27   & 4.53 $\pm$ 0.19    &3.71$\pm$ 0.36 & ...    & 2.10 $\pm$ 0.06 &1.01 $\pm$ 0.09 & 0.49 $\pm$ 0.09& ... \\
     %5.34 $\pm$ 10.1 \\
     %3.83 $\pm$ 34.7
    \hline
\end{tabular}
\end{adjustwidth}
\caption{This table summarizes the fitted parameters of the Band function for different observations, including the spectra indices of lower and higher energy band, $\beta_{1}$ and $\beta_2$, and $E_0$.}
\label{tab:fitting_parameter}
\end{table}

\begin{figure}[ht]
    \centering
    \includegraphics[width =0.9\textwidth]{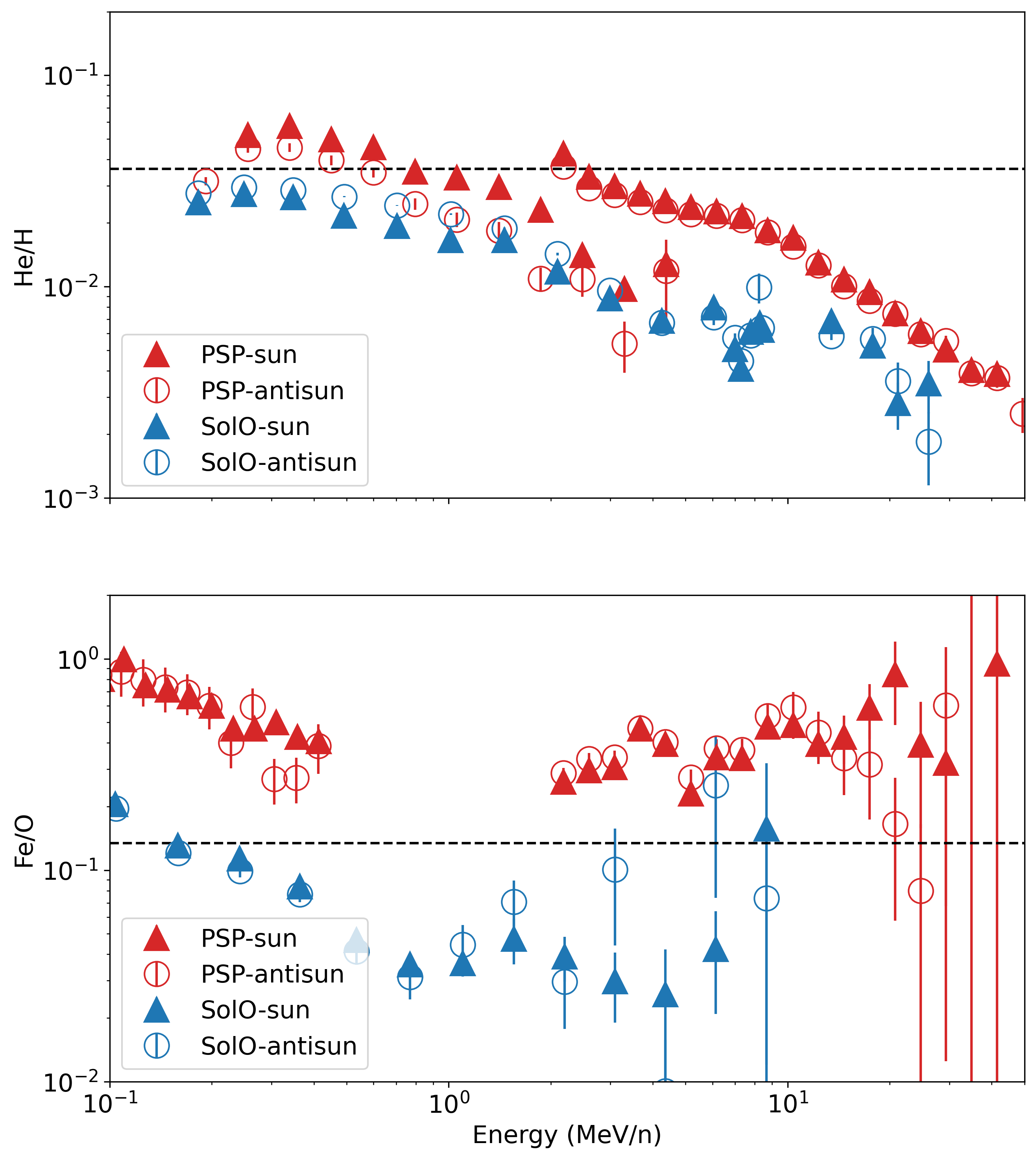}
    \caption{The Fe/O ratio and $^4$He/H ratio measured by PSP and SolO in both sunward and anti-sunward directions as a function of energy. The dashed lines in two panels indicate the averaged values of the He/H ratio (0.036) over 1 - 5 MeV/nuc and the Fe/O ratio (0.134) over 5 - 12 MeV/nuc for large SEP events \citep{Reames1995AdSpR, Reames2017SoPh}.}
    \label{fig:ratio}
\end{figure}

%The key information report from this event, is that this event is the rare case iron rich event in the solar cycle 25, regarding the several years of solar eruption and compared to the previous solar cycle.

Fig.~\ref{fig:ratio} illustrates the energy dependence of the He/H ratio in the top panel and the Fe/O ratio in the bottom panel, for both spacecraft as well as for sunward and anti-sunward views. These ratios are derived from the event-integrated spectra analyzed in Fig.~\ref{fig:spectra}. 
%\ul{Specifically, all the EPiLo fluence from the sunward direction are multiplied by a factor of 4. For protons, only the channels between 0.1 MeV/n and 1 MeV/nuc are utilized, leaving the gap between EpiLo and EpiHi at 1 MeV/nuc.}
The He/H ratios exhibit a strong energy dependence decreasing with increasing energy up to 50 MeV/nuc for both spacecraft and both viewing directions. In contrast, the Fe/O ratio is roughly constant between 0.5 and 10 MeV/nuc for both SolO and PSP, with some increase in the ratio toward lower energies, similar to that reported previously by \cite{Desai2006ApJ}. The sunward and anti-sunward directional ratios agree reasonably well at both spacecraft, although some differences in the He/H ratios may exist below about 1 MeV/nuc (particularly at PSP). 

Particularly notable is the difference in the mean Fe/O ratio observed at PSP compared to SolO.  The mean Fe/O ratio between 0.1 MeV/nuc and 10 MeV/nuc is $\sim$0.48 at PSP, about 5.8 times higher than that at SolO, where the mean ratio is approximately 0.08. Compared with the average Fe/O ratio of 0.134 at about 2 MeV/nuc \citep{reames1998SSRv, reames_heavyelement_2004}  (indicated by the dashed line in the bottom panel of Fig.~\ref{fig:ratio}) this SEP event is Fe-rich at PSP but not at SolO, where the event is Fe-poor. 
Additionally, in the 0.1 to 30 MeV range, the He/H ratio at PSP is approximately 2.2 times higher than it is at SolO. Both of their mean ratios are smaller than the mean He/H ratio of many large and gradual events at $\sim$1- 5 MeV/nuc \citep{Reames1995AdSpR, Reames2017SoPh}, as indicated by the dashed line in the top panel of Fig.~\ref{fig:ratio}.

Note that the uncertainties we considered here are only statistical uncertainties. The systematic uncertainties of those high energy particle telescopes are neglected (generally they are less than a few percent, private communication with the SolO/EPD and PSP/\ISOIS\ team) as other event-specific variations dominate, including differences in the FOVs of the individual telescopes and temporal evolution of the anisotropies, which are difficult to quantify in a single set of uncertainties.

%that for the uncertainty of ratio we give are mainly the statistic uncertainties. The systematic uncertainties are also another factor need to be considered and which could be significant depend on the energy and the instrument. Luckily for the EPiHI, the systematic uncertaintis are element independent, hence the ratio of two elements should be clear out of the systematic uncertainty. The EPILO might be not independent. The number are ***. 
%For SolO, the Fe/O ratio are calculated using SolO/SIS, which like EPIHI whose systematic uncertainties are element independent. Regarding the H/He ratio, we using different instrument, SIS and EPT/HET. They might have different systematic uncertainties. However, even they have the different uncertainties, the systematic uncertainties are still smaller thanthe statistic uncertainty most of time. Hence the systematic uncertainty won't affect our result.

\subsection{Composition variation with time}
\begin{figure}
    \centering
   \includegraphics[width = 0.9\textwidth]{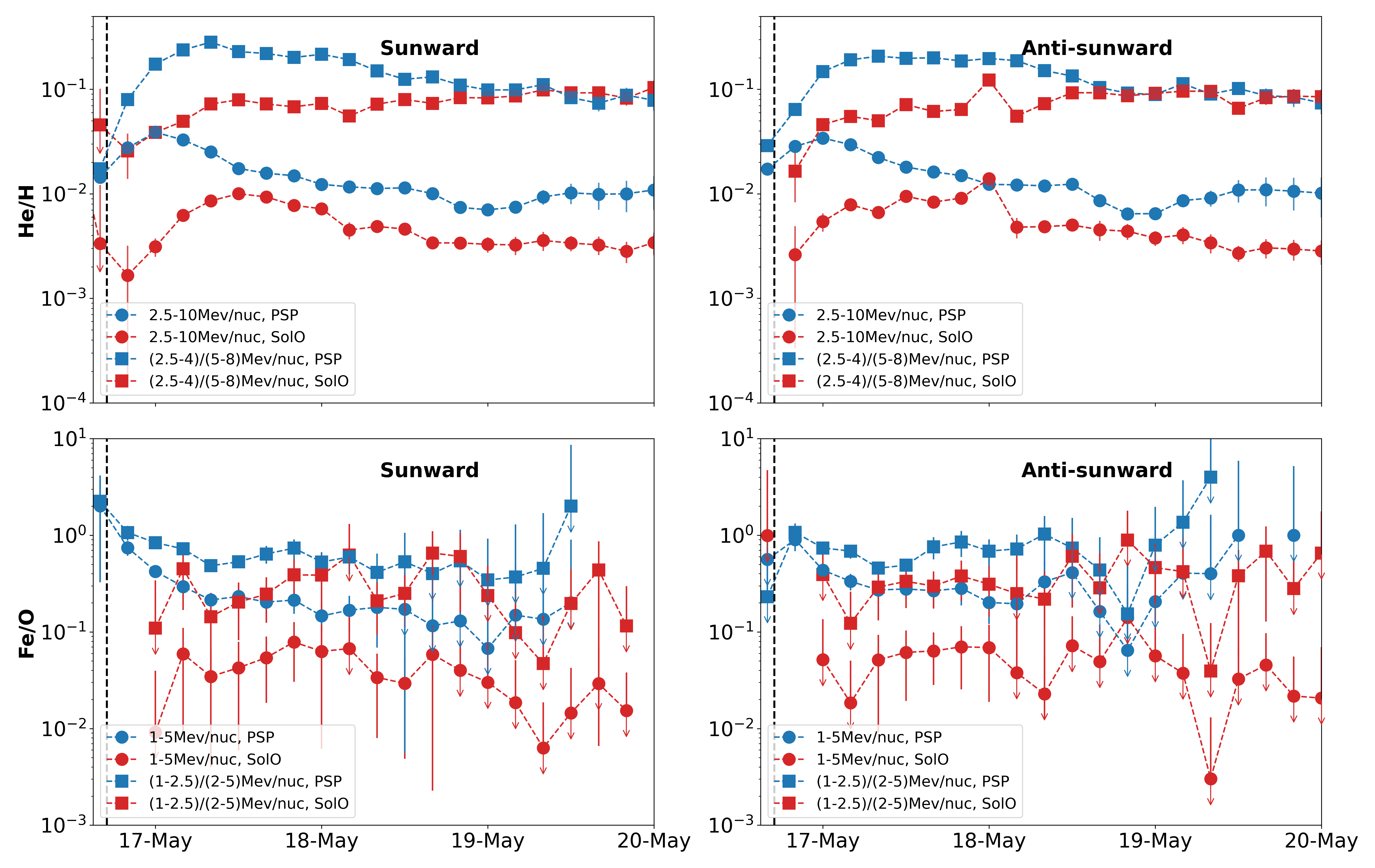}
    \caption{Temporal variation of the Fe/O ratios and He/H ratios measured by PSP and SolO from the sunward and anti-sunward directions.  Ratios calculated at common energies and using the scaled energy are shown, see text for details. The statistical error bars of Fe/O are significantly below the y-axis scale after May 18 due to their lower counting statistics. To simplify the plot, we use downward arrows to indicate the lower limits of uncertainties.}
    \label{fig:temporal}
\end{figure}

\cite{MasonInterplantary2012ApJ, Mason2014ASPC} have demonstrated that a charge-to-mass ratio (Q/M) dependent transport effect could cause a significant increase in the Fe/O ratio during the early phase of an SEP event. Compared to O, Fe has a lower Q/M value, which leads to a larger gyroradius, weaker pitch angle scattering, and a longer mean free path when traveling in interplanetary space. As a result, Fe of the same energy as O can arrive earlier and result in a high Fe/O ratio at the start of the event which decreases with time as the O ions arrive. 

%, we derived the ratio for elements of the same rigidity, as the red and green lines illustrate. After the correction, the time variation of PSP Fe/O ratio at a few MeV/nuc energy channels becomes more constant during the rising phase, in contrast to the original data. 
To examine this concept for our event, we present the temporal evolution of Fe/O and He/H ratios between 1 MeV/nuc and 10 MeV/nuc in Fig.~\ref{fig:temporal}. The ratios are averaged every four hours. The left two panels display the ratio measured by the sunward-facing telescopes of SolO and PSP, whereas the right two panels are from the anti-sunward telescopes. Fig.~\ref{fig:temporal} contains two types of ratios, one where the ratio is calculated from the same energy channels of the two elements. The results are shown as blue circle (PSP) and red circles (SolO) in all panels.  As shown in the bottom left panel of Fig.~\ref{fig:temporal}, at the onset of the SEP event, the sunward Fe/O ratio of PSP shown in blue drops from $\sim$1 to a value closer to the averaged SEP value (0.134) but still enhanced, followed by a period of relative stability. The decrease in the anti-sunward ratio is significantly less, remaining elevated throughout.  

The He/H ratio in the top panel has a similar but more prolonged decrease after an initial increase at the start of the event. Unlike PSP, SolO observed distinct behavior for the Fe/O and He/H ratios. The Fe/O ratio of SolO did not exhibit an initial decline for the energy range from 1 MeV/nuc to 5 MeV/nuc, although the uncertainties are significantly larger than at PSP. The He/H ratio first exhibits a more extended increase before the decrease begins compared to PSP, and the peak ratio is lower than that of PSP.
More importantly, even after the initial phase, the He/H and Fe/O ratios at PSP are still higher than those at SolO. Hence, the difference in the average ratios between SolO and PSP we derived from the fluence spectra is not only due to the large differences during the initial phase, but also present in the decay phase. 
%Unfortunately, in this event, we could not find any perfect energy channels from SolO and PSP that both can provide a Fe/O ratio curve with smaller uncertainty. 

%The Fe/O ratio in the energy of 80 - 130 keV/nuc from SIS (not shown) have the initial decline, like the blue line in the bottom left panel.

Following the methodology of \cite{MasonInterplantary2012ApJ}, to correct for the transport effect of particles at the start of the event, we derive the abundance ratios et energies where both particle species have equal diffusion coefficients. In our calculation, the energies of Fe, $E_{Fe}$, and He, $E_{He}$ with the same diffusion coefficients as O and H, are half of the energies of O, $E_O$ and H, $E_H$. This scaling energy factor is estimated from Eq. 1 of \cite{MasonInterplantary2012ApJ}, where the typical charge state of the SEP ions of O of 6.8 and Fe of 11.6 and the typical rigidity dependence of the scattering mean free path are used, which is estimated from the power spectrum of magnetic turbulence (See \cite{MasonInterplantary2012ApJ} for more details). The corresponding ratios from PSP and SolO are shown as blue and red squares, respectively. The time profiles of the PSP sunward and anti-sunward Fe/O ratios at the scaled energies are somewhat flatter at the start of the event, but the Fe/O ratios at SolO have similar temporal trends as the ratios calculated at the same energy.
During the first half of the event, the flattened ratios at same diffusion coefficient observed by PSP are still higher than those observed by SolO, indicating distinct particle populations at both locations. This difference is particularly manifested in the He/H ratios, as shown by the blue and red squares in the top two panels. The PSP He/H ratio is nearly an order of magnitude higher than the SolO ratio at the start of the event. As the event evolves, this discrepancy gradually decreases and eventually vanishes by the end of the event.
Regarding the Fe/O ratios calculated at the scaled energies, a slight difference is also observed, in particular at the beginning of the event; however, we refrain from overinterpreting the results because of the relatively large uncertainties associated with the measurements from both spacecraft.

%\TODO{I need to go through the paper to decide the following}
%\st{
%The corresponding changes of SolO Fe/O ratio with energy above 1 MeV/nuc are unclear due to significant uncertainty. Furthermore, the initial drop feature of the ratio between Fe and O with the energy of 0.08 - 0.12 MeV/nuc from SIS which has more counting statistics is no longer evident, after considering the scaled energy of O. Nevertheless, during the rising phase of the SEP event, same rigidity ratio of PSP is still higher than that of SolO, in both He/H and Fe/O ratio.}

%The decrease at SolO He/H shows different some different, compare to Fe/O on explanation is the wave amply the intensity. cause the first increase of the He/H ratio.

%\begin{equatin}
%    \frac{E_a}{E_b} = [\frac{(Q/M)_a}{(Q/M)_b}]^{2\alpha/(\alpha+1)}
%\end{equation}

%The shock arrival at 5.18 together with the decrease of the ratio in both cases.

%Fe/O: Fe enhancement during the SEP onset (Mason 2014, 2012)
%    Fe/O variation ....

%Q/M is the particle change-to-mass ratio. $\alpha$ is the rigidity dependence of the scattering mean free path which depends on the power spectrum of magnetic turbulence; for the typical SEPs ionization states and $\alpha$=1, the scaling factor between O and Fe is $\sim$ 2 

%Both SolO and PSP, He/H ratios increase at the start of the event. The increase of SolO last longer than PSP; PSP last 8 hours, SolO last at least 16 hours for Sunward. While 

\section{Discussion and Summary}

The key result for the SEP event on 16 May 2023 is that it is Fe-rich at PSP but Fe-poor at SolO. So far this is the only reported Fe-rich  event observed at PSP in solar cycle 25 aside from the well-known, first Ground Level Enhancement (GLE) event of this cycle on 28 Oct 2021.  That event was found to be Fe-rich at multiple spacecraft \citep{Cohen2022cosp, Guo2023GeoRL, Kouloumvakos2024AA}.
%the second . 
During our event, PSP and SolO formed a special constellation being at the same radial distance from the Sun but having a 60$^\circ$ separation in longitude, providing a unique opportunity to examine the longitudinal  variation of the SEP event composition. Both spacecraft's magnetic footpoints are located west of the flare, but PSP's footpoint is much closer to the flare ($\sim$18$^\circ$) than SolO's ($\sim$80$^\circ$). %The magnetic foot point of the SolO is about 
%which have less than anticipated iron rich even than the previous solar cycles (Christina, private communication). 
By combining data from the various instruments on PSP and SolO, we obtained event-integrated fluence spectra of H, He, O, and Fe between 0.1 MeV/nuc and a few tens of MeV/nuc in both the sunward and anti-sunward viewing directions. All the spectra exhibited a typical double power law shape, and were well fit by a Band function.  From these spectra we calculated He/H and Fe/O abundance ratios.  Of particular note is the average Fe/O ratio between 0.1 and 10 MeV/nuc observed on PSP was about 5.8 times higher than that on SolO.

%The variability of ion abundances, specifically the higher Fe/O ratio during the initial phase can be accounted for by many factors including different acceleration processes, the interplanetary transport effect, and the geometry of CME-driven shocks that accelerate suprathermal seed populations \cite{Cane2003, 2006, Tylka 2005, Mason 1999, 2014?, Cohen 2014}. Of them, 

The strong longitudinal dependence of the Fe/O ratio observed in this event seems consistent with a scenario in which there is a direct contribution of flare-accelerated material to the SEP event. Studies by \cite{Cane2003GeoRL, Cane2006JGR} found that higher Fe/O ratio events are usually associated with well-connected western events that have larger flare contributions and are accompanied by weaker CME-driven shocks. Conversely, normal coronal compositions are measured in the remote regions far from the field line connected to the source region. Thus, the Fe/O ratio should be higher in a narrow region well connected to the source region. In this event, the magnetic footpoint of PSP is within 20$^\circ$ of the flare region, and the Fe/O ratio was much higher than that for SolO whose footpoint was 80$^\circ$ west of the flare. We have fitted the Fe/O ratios from the two spacecraft with a Gaussian distribution assuming a center at the flare location similar to that done by \cite{Cohen2017ApJ}. The angular width in this event is narrower than that of the longitudinal distribution of the Fe/O ratio SEP events in solar cycles 23 and 24, as derived in \cite{Cohen2014ApJ, Cohen2017ApJ}. This result suggests that the source of the Fe-rich component originated from a limited region on the solar surface. We also examined the $^3$He observations at both locations, as flare-acceleration processes are often accompanied by enhancements of $^3$He/$^4$He as well as that of Fe/O. However, neither PSP/LET nor SolO/SIS detected significant $^3$He during this event. The hourly He mass histograms from PSP/LET show that $^3$He count rates rarely exceed the background expected due to spillover from $^4$He, except during the first few hours at the start of the event. Even then, the total counts of $^3$He in each hour is less than 10 with a $^3$He to $^4$He ratio between 0.10 and 0.15. 
%I noticed a pre-event background of He4 particle which might related to the solar activities happened on the solar surface. So it might be an indication of the existence of flare seed population between the PSP and Sun, which is fresh and ne

%\TODO{is there relationship between Fe/O and He3/He4 ratio we could give an estimate of the expected He3/He4 ratio. ONe figure.}
Moreover, the enhancement of $^3$He is not strictly correlated with a high Fe/O ratio. \cite{Cohen2014ApJ} indicate that the SEP event on 11 April 2013 observed by STEREO and ACE, despite the fact that their magnetic footpoints were on the opposite sides of the flare rather than well connected to the flare, had a high Fe/O ratio at both locations. However, almost no $^3$He  was measured at either spacecraft, and only observational upper limits of $\rm ^3He/^4He$ of a few percent were derived. \cite{Cohen2014ApJ} suggest that, although the SEP event was Fe-rich at both spacecraft, the lack of a longitudinal variation did not support the direct flare contribution scenario. 

\begin{figure}
    \centering
    \includegraphics[width = \textwidth]{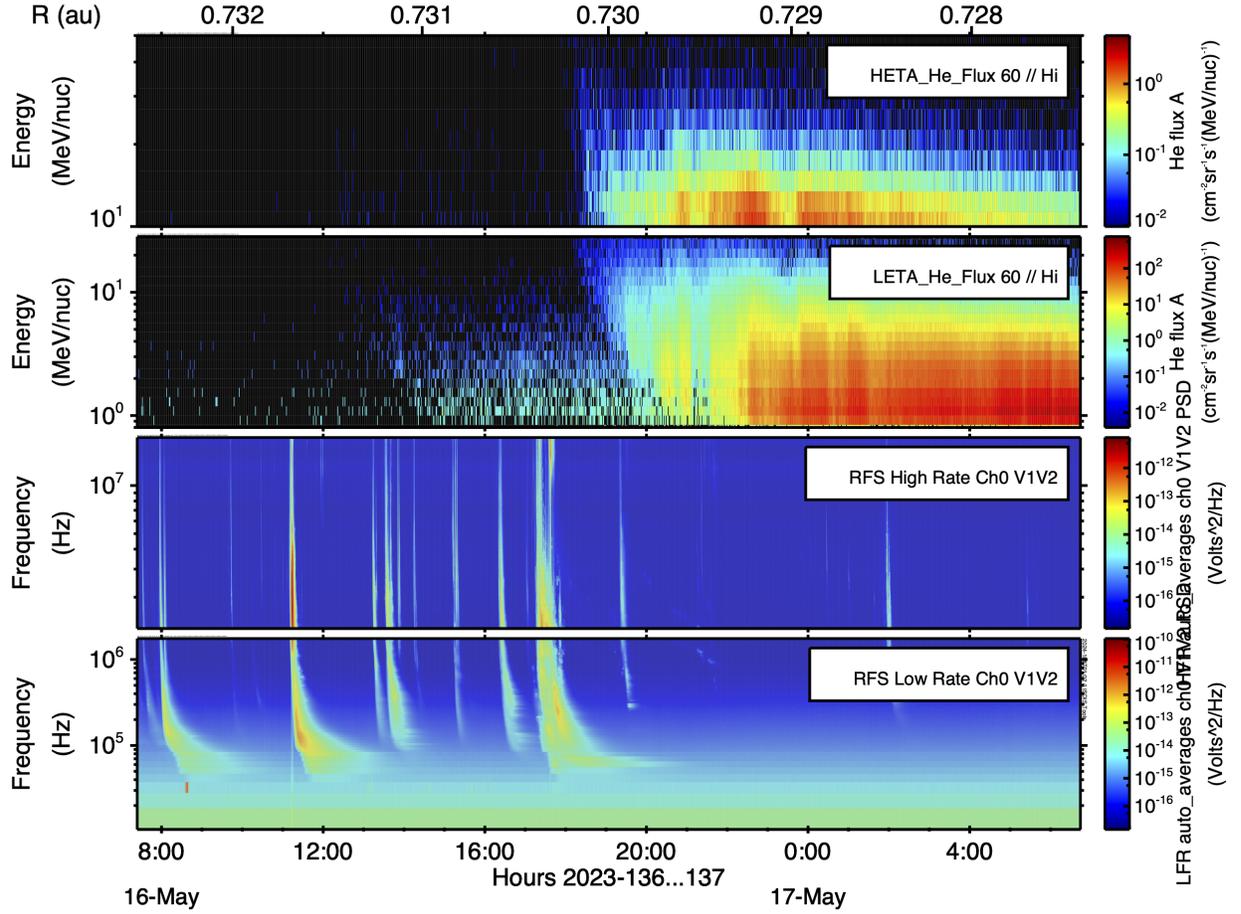}
    \caption{Top two panels: $^4$He observation on PSP/LET and PSP/HET; Bottom two panels: The radio observation on PSP/FIELD.}
    \label{Fig:radio}
\end{figure}

The second possible mechanism to explain Fe-rich events involves Fe-rich seed populations from flare suprathermal remnants %with higher energy than the solar wind suprathermal particles,
being accelerated by a quasi-perpendicular shock \citep{Tylka2005ApJ, Tylka2006ApJ}. Since the injection energy threshold of quasi-perpendicular shock is expected to be higher than for a parallel shock, more suprathermals and fewer bulk solar wind particles will be accelerated to SEP energies. In this scenario,  in general, no or little longitudinal dependence is expected in the Fe/O ratio. If there is any, it would be determined by the distribution of the suprathermal seed particles near the Sun and the geometry of the shock when it encounters those seed particles. Fig.~\ref{Fig:radio} presents the He dynamic spectrogram from PSP/LET and PSP/HET between May 16 at 8:00 and May 17 at 6:00, covering a few hours before the event and the initial phase of the SEP event. Interestingly, the pre-event intensity of He at energies below 10 MeV/nuc increases concurrently with the occurrence of multiple type III radio bursts observed by PSP/FIELDS/RFS, as shown in the bottom two panels of Fig.~\ref{Fig:radio}. This might suggest the existence of flare-related remnant particles between PSP and the Sun, which can be further accelerated by the CME-driven shock occurring later. SolO/EUI and STIX observations indicate that the AR13296 is likely the source of these seed particles.
However, this increase in He was not suitable for VDA analysis so the solar release time is unknown.
%this increase in He did not exhibit any clear velocity dispersion that could be used to determine a solar release time. 
%\st{Additionally, we cannot determine the location of the solar source of these type III eruptions. SolO/EUI observations reveal multiple active regions on the back side of the Sun that have flaring activity during the pre-event period.}

%Before the flare eruption we are focusing on, there was an M9.6 class flare situated on the eastern limb of the Sun, closer to SolO's magnetic footpoint on the solar surface. The corresponding SXR light curve peaks at 16:42, based on GOES and SolO/STIX observations. This eruption is accompanied by a large, bright filament and a slow-moving CME. The CME front can be seen from the bottom-left corner of Panel (C) in Fig.~\ref{fig:Overview}. However, it is unlikely to have any effect on the event seen by PSP or SolO.

Generally, the shock is expected to be more quasi-perpendicular at the flanks of the CME than at the nose, where it is typically quasi-parallel. Thus, it could be argued that the SEP event is more likely to be Fe-rich at an observer near the flank of an outwardly-propagating CME, which is not where PSP is connected to at the time of the particle release in this event.  However, without a detailed simulation of the shock propagation in space, the exact shock geometry and how its evolution affects the composition seen at different locations remains unclear. Nevertheless, the possibility of the CME-driven shock generating the higher Fe/O ratio observed at PSP cannot be entirely ruled out but the flare contribution scenario appears to be a simpler, and therefore more likely cause.

Another SEP event in this solar cycle when PSP and SolO were situated at a similar radial distance but separated in longitude, is the first widespread SEP event of this solar cycle. That SEP event occurred on 29 November 2020 and was recorded by more than 6 observers, including PSP and SolO, with a longitudinal spread of more than 230$^o$ \citep{Kolhoff2021AA, Cohen2021AA, Mason2021AA}. During the event, the radial distance of SolO was 0.88 au and PSP was 0.81 au. The magnetic foot of SolO was situated about 89$^o$ east of the flare, while PSP was closer with a separation of 48$^o$ to the west. Neither spacecraft was well connected to the flare and the heavy ion abundance analysis \citep{Mason2021AA} shows that this event had a lower than average Fe/O ratio at energies above a few MeV/nuc with the ratio decreasing rapidly with increasing energy evident in  PSP, ACE, and STEREO-A observations. In contrast, the 28 Oct 2021 event was observed to have higher Fe/O ratios than the average at both ACE (1 au), PSP (0.62 au), and STEREO-A (0.96 au) above few MeV/nuc \citep{Kouloumvakos2024AA,Cohen2022cosp} when those spacecraft were located within $\sim$60$^o$ of each other. Both events have very different longitudinal dependencies compared with the May 16, 2023 event, but also do not have optimal configurations of spacecraft location relative to the flaring region for testing the two scenarios proposed to explain Fe-rich SEP events. 

Since the start of solar cycle 25 at the end of 2019, the Sun has been more active with the number of large solar eruptions and large SEP events continuing to increase. The occurrence of Fe-rich SEP events appears to be less frequent than in solar cycle 23, providing fewer opportunities for the kind of study presented here. As the SolO and PSP missions progress, we anticipate more chances to observe SEP events in closer proximity to the Sun. These observations will aid in understanding the cause of the Fe-rich component of SEP events  and potentially address the different characteristics of SEP events over different solar cycles.

\begin{acknowledgments}
We acknowledge the contribution of the Solar Orbiter and Parker Solar  Probe mission team, especially the contribution of the Energetic Particle Detector and the \ISOIS \ instrument team when analyzing the data. Parker Solar Probe was designed, built, and is now operated
by the Johns Hopkins Applied Physics Laboratory (JHU/APL)
as part of NASA’s Living with a Star (LWS) program (contract
NNN06AA01C). Support from the LWS management and
technical team has played a critical role in the success of the
Parker Solar Probe mission. We thank the scientists and
engineers whose technical contributions prelaunch have made
the \ISOIS \ instruments such a success. 
Solar Orbiter post-launch work at JHU/APL and SwRI is supported by NASA contract 80MSFC19F0002. EPD is supported by the German Space Agency, DLR, under grant 50OT2002 and the Spanish MCIU/AEI Project PID2019-104863RBI00/AEI/10.13039/501100011033. A.K. acknowledges financial support from NASA NNN06AA01C (PSP EPI-Lo, SO-SIS Phase-E) contract.  C.M.S.C. acknowledges additional partial funding from NASA grants 80NSSC22K0893, 80NSSC21K1327, 80NSSC20K1815, and 80NSSC19K0067.

\end{acknowledgments}

\bibliography{main}{}
\bibliographystyle{aasjournal}



\end{document}